\documentclass[journal=jacsat,manuscript=article]{achemso}

\usepackage{physics}
\usepackage[T1]{fontenc}
\usepackage[utf8]{inputenc}
\usepackage{amssymb}

\usepackage{fix-cm}

\usepackage{amsmath} 
\usepackage{xcolor} 
\usepackage{soul}   

\author{Krystyna Kolwas}
\email{Krystyna.Kolwas@ifpan.edu.pl}
\phone{+48 22 116 2261}
\fax{+48 22 843 0926}
\affiliation[Institute of Physics, Polish Academy of Sciences]{Institute of Physics, Polish Academy of Sciences,\\Al. Lotników 32/46, 02-668 Warsaw, Poland}

\title[An \textsf{achemso} demo]
  {Self-Quenching Effect of the Decay of Localized Surface Plasmons: Classical and Quantum Perspectives\footnote{A footnote for the title}}

\abbreviations{LSP, PQP, DR, LDOS, TM, TE, CQED, PCQED, NP, MNP, QNM, FGR}

\begin{document}

\begin{tocentry}
    \includegraphics[width=7.6cm]{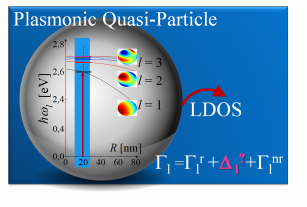}
   \end{tocentry}

\begin{abstract}
This study presents a self-consistent, quantum-informed model for the decay dynamics of localized surface plasmons (LSPs) in spherical metal nanoparticles (NPs), described as plasmonic quasi-particles (PQPs). By bridging classical electrodynamics description for quasi-normal modes (retardation effects included) with a quantum emitter perspective, this framework provides an analytically tractable description of the damping of the dissipative confined plasmonic systems. In addition to its significance for emission control, the model emphasizes the bosonic characteristics of plasmonic quasi-particles, which are coherent many-electron excitations of the states of quasi-normal modes. 
Unlike conventional cavity quantum electrodynamics (CQED), where the emitter and cavity exist as separate systems, a plasmonic quasi-particle functions as a quantum emitter embedded within a self-created resonant near-field nano-cavity of confined radial fields, sharing the spectral characteristics of the surface transverse-magnetic (TM) modes, which include nonradiative damping effects resulting from, e.g., ohmic losses in a metal. This work extends Fermi's Golden Rule to include the coupling between the emission process and the self-generated cavity impact. The derived self-consistent formulation offers analytical expressions for the total damping rates, which demonstrate a size-dependent suppression displayed in higher multipolarity modes attributed to the impact of the self-quenching effect resulting from the coaction of radiative and non-radiative channels.
\end{abstract}

\section{Introduction}

Nanometallic (plasmonic) structures exhibit a remarkable ability to confine and concentrate electromagnetic (EM) field energy within subwavelength volumes, making them essential components of modern nanophotonics \cite{barnes2003surface,schuller2010plasmonics,novotny2012principles,alekseeva2019single,nam2019chemical}. Their ability to create highly localized and intense optical fields has prompted a fundamental reevaluation of light–matter interactions at the nanoscale. The field of plasmonics provides a rich platform for manipulating light at nanometer scales, with applications ranging from nanophotonic lasers \cite{gaio2019nanophotonic} and optical amplifiers \cite{qin2024quantum} to metamaterials \cite{cai2010optical}, biochemical sensing \cite{aslan2005plasmon,jain2007review,shalabney2011sensitivity}, and optical nanoantennas \cite{monticone2017optical}.

Recent research has increasingly focused on the quantum aspects of localized surface plasmons (LSPs) \cite{jacob2011plasmonics,van2012spontaneous,scholl2012quantum,bozhevolnyi2017plasmonics,kolwas2019decay}. In conventional cavity quantum electrodynamics (CQED), the spontaneous emission of a quantum emitter can be strongly modified by placing it inside an optical cavity whose spectral and spatial characteristics alter the vacuum fluctuations of the electromagnetic field \cite{haroche1989cavity,walther2006cavity,guzatov2012plasmonic,hugall2018plasmonic,franke2019quantization}. 

In plasmonic implementations of CQED (PCQED), nanoscale metallic cavities confine electromagnetic fields to extremely small effective volumes while maintaining significant field intensities. In such systems, the interaction dynamics between the cavity field and an atom- or molecule-like emitter is determined not only by the cavity quality factor but also by the strongly reduced effective mode volume \cite{kristensen2014modes,ginzburg2016cavity,hugall2018plasmonic,wei2021plasmon}. The resulting modification of the local density of optical states (LDOS) typically leads to substantial enhancement or suppression of spontaneous emission rates via the Purcell effect.

However, conventional approaches typically treat the quantum emitter and the cavity as distinct, independent subsystems. The present work addresses a fundamentally different and highly original physical scenario: the decay dynamics of a plasmonic excitation that acts simultaneously as the emitter and as the source of its own structured electromagnetic environment. Here, the plasmon itself is described as a plasmonic quasi-particle (PQP) — a collective, bosonic many-electron excitation embedded within a self-created resonant near-field nanocavity of confined radial fields.

A cornerstone of standard decay models of atom-like emitters is the assumption of additivity — namely, that the total damping rate of an excitation can be expressed as the sum of independent radiative and nonradiative decay channels. In the case of localized surface plasmons (LSP) excitations, we show that this assumption does not generally hold. Within the self-generated electromagnetic environment of a plasmonic nanoparticle, intrinsic nonradiative (ohmic) losses modify the local density of optical states (LDOS) of the confined field. This dynamic coupling introduces a feedback mechanism that alters the effective radiative decay rate of the plasmonic excitation. As a result, the radiative and nonradiative decay channels become intrinsically coupled and cannot, in general, be treated as independent contributions.

To capture this effect, the theoretical framework developed here bridges classical Maxwell electrodynamics with a quantum description of plasmonic excitations. The classical dispersion relations for surface-bound transverse-magnetic modes, with surface collisions and retardation effects fully included, provide the eigenfrequencies and damping rates of the quasi-normal modes supported by spherical nanoparticles \cite{fuchs1992basic,kolwas1996optical,kolwas2006smallest}. These classical results serve as the foundation for constructing a discrete quantum description of LSP excitations and for analyzing their decay dynamics within a self-consistent emitter–cavity system.

On this basis, we derive a self-consistent extension of Fermi’s Golden Rule for plasmonic quasi-particles that incorporates the coupling between the emission process and the self-generated cavity impact. This formulation yields an analytical expression for the total damping rates, revealing a size-dependent reduction in the total damping rate, particularly prominent in larger size ranges and higher multipolarity modes. This intrinsic suppression of radiative emission—referred to here as the self-quenching effect — arises directly from the nonadditive coaction of radiative and nonradiative channels.

Ultimately, the resulting framework provides a unified perspective that connects classical electrodynamics to the quantum bosonic characteristics of plasmonic quasi-particles, offering novel, analytically tractable insights into the fundamentally coupled decay dynamics of localized surface plasmons.

\section{Classical Maxwell Foundations}

In classical electrodynamics, the response of a spherical nanoparticle (NP) of radius \( R \) to an external electromagnetic (EM) wave is described by solving Maxwell’s equations with boundary conditions at the NP surface. The EM fields inside and outside the NP are typically expanded using Lorenz–Mie scattering theory as an infinite series of spherical multipole modes with angular momenta \( l = 1, 2, 3, \ldots \) \cite{born1999principles}. The general solution is a sum of transverse magnetic (TM) and transverse electric (TE) modes.

At resonance, the incident EM wave couples efficiently to the curved metal–dielectric interface, generating standing surface waves whose wavelengths scale with \( R \). These waves produce confined radial near-fields and scattered radiation detectable in the far field. LSP resonances are often identified with the maxima in these far-field spectra, which are experimentally observable for NPs of known radii.

However, scattering theory does not explicitly provide the resonance frequencies or damping rates of LSPs. Their dependence on radius, material, and surrounding medium arises from quasinormal modes (QNMs) — complex eigenmode values defined by the dispersion relation (DR) for surface-confined EM waves. These modes are coupled to collective electron oscillations at the metal–dielectric boundary.

\subsection{Dispersion Relation for TM Modes}

The dispersion relation for QNMs associated with LSPs is derived from the source-free Mie formalism \cite{mie1908}, as presented in classical works \cite{stratton2007electromagnetic, born1999principles, ruppin1982electromagnetic, fuchs1992basic}. Scalar solutions to the Helmholtz equation yield two distinct vectorial modes:  TE modes: \( E_r = 0 \),
and TM modes: \( E_r \neq 0 \), which alone couple to surface charge oscillations.

The complex frequencies of QNMs are roots of the dispersion relation for TM modes with angular momentum \( l \), derived from the continuity of EM field components at the interface \( r = R \) \cite{fuchs1992basic,kolwas1996optical,kolwas2006smallest}:

\begin{equation}
\sqrt{\varepsilon_{in}}\,\xi_l' (k_{out} R)\, \psi_l (k_{in} R) 
= \sqrt{\varepsilon_{out}}\,\xi_l (k_{out} R)\, \psi_l' (k_{in} R).
\label{DR}
\end{equation}

Here, \( \varepsilon_{in} \) and \( \varepsilon_{out} > 0 \) are the dielectric functions of the metal and its environment, respectively. The wave numbers are \( k_{in} = \sqrt{\varepsilon_{in}}\,\omega/c \) and \( k_{out} = \sqrt{\varepsilon_{out}}\,\omega/c \); \( \psi_l \) and \( \xi_l \) are Riccati–Bessel functions. 

Unlike the commonly employed driven-propagation spectral framework for planar surface plasmon polaritons (SPPs), in which the dispersion relation yields complex wave vectors at real excitation frequencies, the quasi-normal mode description of LSPs on spherical metallic nanoparticles necessitates the allowance of complex eigenfrequencies. This requirement is analogous to the complex-frequency eigenmode formulation used for planar SPPs \cite{lalanne2018structural}:

\begin{equation}
\Omega_l(R) = \omega_l(R) - i \Gamma_l(R),
\label{roots}
\end{equation}

where \( \omega_l(R) \) is the resonance frequency and \( \Gamma_l(R) \) is the total damping rate of LSP oscillations of QNM TM modes.

The functions \( \omega_l(R) \) and \( \Gamma_l(R) \) describe discrete standing surface fields, their radial confinement, and far-field radiation. TE modes are not coupled to the interface as they lack solutions at \( r = R \), confirming that only TM modes contribute to LSP resonances.

\begin{figure}[ht]
\centering
\includegraphics[width=8.0cm]{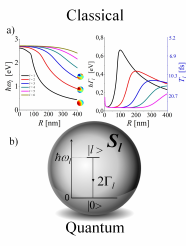}
\caption{Classical and quantum representations of LSP dynamics. 
(a) Resonant frequencies \( \hbar\omega_l(R) \) and damping rates \( \hbar\Gamma_l(R) \) of QNMs as a function of radius \( R \), including retardation effects \cite{kolwas2013damping}. 
(b) Mapping multipolar modes \( l \) to energy levels of a quantum plasmonic quasi-particle (PQP), modeled as independent two-level systems \( S_l \).}
\label{fgr:1}
\end{figure}

Figure~\ref{fgr:1}a illustrates the dependence $\omega_l(R)$ and $\Gamma_l(R)$  for gold nanospheres for consecutive radii (data from \cite{kolwas2013damping}), retardation included). 

Spatial field confinement stems from the negative real part of \( \varepsilon_{in}(\omega) \) (anomalous dispersion) in a dielectric medium (\( \varepsilon_{out} > 0 \)), which causes the radial electric field to decay evanescently. Outside the NP, radial components fall off as \( 1/r^2 \), and tangential components as \( 1/r \) \cite{fuchs1992basic}.

The dispersion-relation approach has been the subject of our extensive studies employing a size-dependent dielectric function  $\varepsilon_{in}(\omega,R)$,  which explicitly accounts for surface-collision effects arising from finite nanoparticle sizes as well as retardation effects in gold and silver nanoparticles embedded in various dielectric environments {\cite{{derkachova2016dielectric, kolwas2017modification, kolwas2020impact}}}. By numerically solving the corresponding equations for a consecutive nanoparticle radii, we obtained the radius-dependent resonance frequencies $\omega_l(R)$ and the associated damping rates $\Gamma_l(R)$ over an exceptionally wide range of particle sizes, from a few nanometers up to several hundred nanometers. Importantly, the size evolution of these quantities is not imposed through predefined scaling relations but emerges directly and self-consistently from the dispersion equation.

This methodology is applicable over an even broader size regime, extending in principle from the nanometer to the micrometer scale, provided that the employed dielectric response function accurately reproduces the metal’s optical properties in the spectral domain relevant to the multipolar LSP oscillation frequencies under consideration. The frequency window within which the dielectric function remains valid determines the admissible range of complex solutions $\omega_l(R) + i\Gamma_l(R)$. However, this constraint lies well outside the size range pertinent to the phenomena analyzed here and to typical experimental implementations.

\subsection{Damping Rates of Multipolar LSP Oscillations vs Radius}

\begin{figure}
\centering
\includegraphics[width=8.0cm]{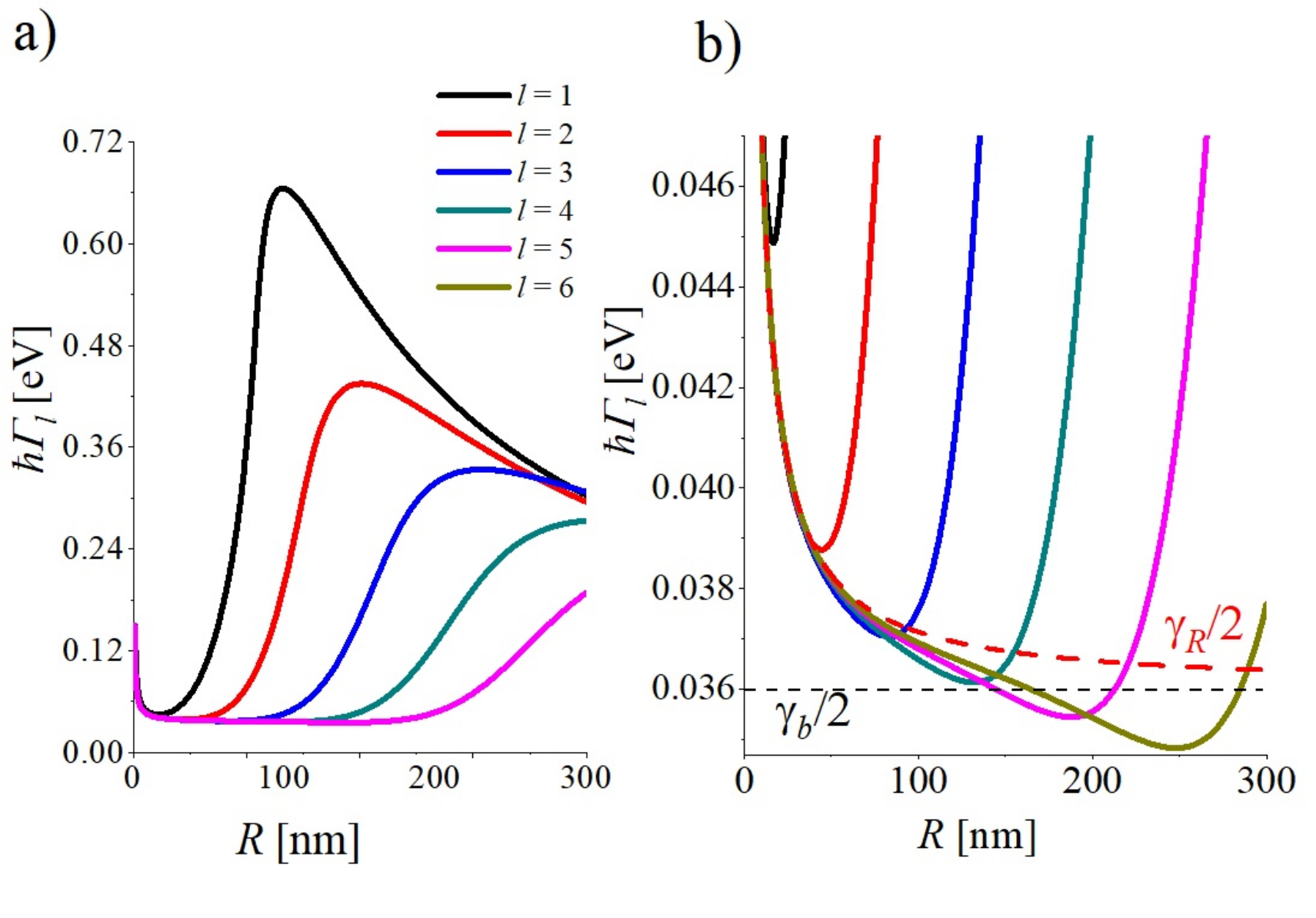}
\caption{(a) Total damping rates $\Gamma_l(R)$ (for $l$ = 1 - 6) calculated for gold NPs from DR (data from {\cite{kolwas2013damping}}). 
(b) Close-up view showing that $\Gamma_l(R) \leq \gamma(R)/2$, suggesting non-additivity of radiative $\Gamma_l^r$ and nonradiative $\gamma(R)/2$ contributions.}
\label{fgr:reduction}
\end{figure}

The total damping rate \( \Gamma_l(R) \) of LSP oscillations (see the example in Fig.~\ref{fgr:reduction}a)) includes both radiative losses and nonradiative processes resulting from collisional damping.  
It is often assumed to be additive:

\begin{equation}
\Gamma_l(R) \simeq \Gamma_l^r(R) + \Gamma^{nr}(R),
\label{G_l}
\end{equation}

where \( \Gamma_l^r(R) \) is the radiative damping rate (from the collisionless model), and \( \Gamma^{nr}(R) \simeq  \Gamma^{col}(R)= \gamma(R)/2 \), if assumed to account for electron collisions, with:

\begin{equation}
\gamma(R) = \gamma_b + \frac{A v_F}{R}.
\label{gamma}
\end{equation}

This effective collision rate appears explicitly in \( \varepsilon_{in} \), incorporating both bulk ohmic losses \( \gamma_b \) and the size-dependent contribution from surface scattering of electrons, where \( v_F \) denotes the Fermi velocity and \( C \approx 1 \) for noble metals such as Au and Ag (e.g., \cite{derkachova2016dielectric, kolwas2020impact}). The data from \cite{derkachova2016dielectric} (Figure~\ref{fgr:reduction}) also phenomenologically incorporate the influence of interband transitions on the polarizability. However, the corresponding transition rate is not an explicit parameter in the model of the dielectric function.

In the small size ranges of noble-metal spherical NP's that are experimentally available ($R\lesssim 80$nm), nonradiative damping with the rate $\Gamma^{nr}(R) = \gamma(R)/2$ dominates (Fig.~\ref{fgr:reduction}). Dipolar mode (\( l = 1 \)) in this range is weakly radiating, with a small contribution of \( \Gamma_l^r(R) \) to \( \Gamma_l(R) \) ("low-radiative modes").  Higher-order modes are mostly nonradiative in such a size range ("dark modes").

As \( R \) increases, radiative losses grow, eventually dominating total damping ("bright" or radiant modes).   
Interestingly, numerical results reveal that for some size ranges: (see Fig.~\ref{fgr:reduction}b):

\begin{equation}
\Gamma_l(R) \leq \Gamma^{nr} = \frac{\gamma(R)}{2},
\label{Gamma_nr_corr}
\end{equation}

indicating that a simple additive model is insufficient, as in general: $\Gamma_l(R) \leq \Gamma_l^r(R)+ \Gamma^{nr}(R)$.
To reconcile this, we define a correction term:

\begin{equation}
\Delta_l(R) := \Gamma_l(R) - \left( \Gamma_l^r(R) + \Gamma^{nr} \right) \leq 0,
\label{Delta_l}
\end{equation}

which may stem from a suppression of the nonradiative damping \cite{sonnichsen2002drastic,kolwas2013damping}. In particular, Sonnichsen et al.~\cite{sonnichsen2002drastic} attributed the suppression of nonradiative damping primarily to a reduction in interband transition processes.

However, the arguments based on classical Maxwell electrodynamics do not fully account for the reasons for these reducing corrections to the total damping rate of the radiative modes:
\begin{equation}
\Gamma_l(R) = \Gamma_l^{r}(R) + \Delta_l(R)+ \Gamma^{nr}(R) \leq \Gamma_l^r(R) + \Gamma^{nr}(R).
\label{Gamma_l}
\end{equation}

A tentative interpretation of the hypothesis, that introduces the possible co-dependence of the radiative and nonradiative damping channels, and examines the consequences for understanding the intrinsic radiative decay processes of quantum plasmonic emitters, is the subject of the Sections that follow.

\section{Foundations for Describing LSP Excitations as Quantum Plasmonic Quasi-Particles}

The concept of modeling localized surface plasmon (LSP) excitations as quasi-particles in the excited state arises from their resonant behavior followed by relaxation, observable through the spectral properties of scattered or absorbed light. This approach establishes a bridge between the classical electromagnetic description based on Maxwell’s equations and a quantum-mechanical framework where plasmonic excitations are treated as quantized states of a collective many-electron system.

The solutions of the dispersion relation for transverse-magnetic (TM) modes form the classical foundation for the quantum framework. In this view, plasmonic excitations can be conceptualized as discrete, quantized many-electron states with energies \( \hbar\omega_l \) (see Figure~\ref{fgr:1}b).  

Each multipolar mode \( l \) corresponds to a discrete energy level of an excited quantum plasmonic quasi-particle (PQP), modeled as a set of uncoupled two-level subsystems \( S_l \) undergoes transient decay to the ground state \( |0\rangle \) after excitation.

Unlike atomic energy levels, the states of a plasmonic quasi-particle \( \hbar \omega_l(R) \) originate from the collective oscillations of conduction electrons. The decay of these states includes both population relaxation (energy dissipation) and coherence decay (dephasing). Damping occurs via radiative emission as well as nonradiative mechanisms such as ohmic heating caused by electron–electron and electron–surface scattering within the confined NP volume. Nonradiative losses dominate in small nanoparticles, whereas radiative damping becomes the leading process as the particle size increases.

\subsection{Closed System}

In conventional CQED, models such as the Jaynes-Cummings Hamiltonian describe the interaction between a two-level atom and a single quantized mode of an external cavity field \cite{jaynes1963comparison}. This model assumes a closed system that conserves the number of quanta. 

\begin{figure}
\centering
  \includegraphics[width=8.0cm]{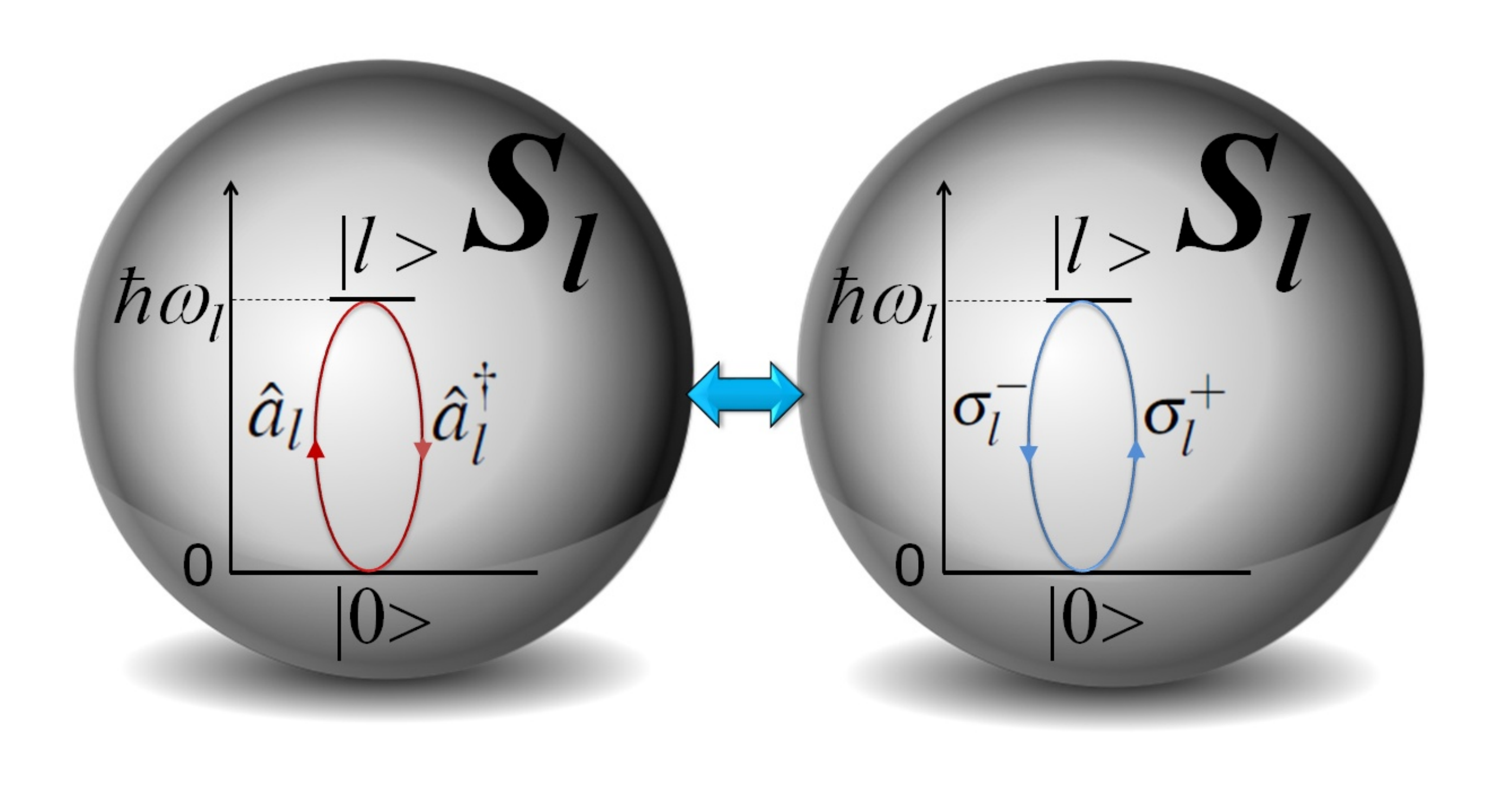}
  \caption{Conceptual link between the quantum closed cavity-mode model and the plasmonic quasi-particle (PQP) picture. Each TM surface mode of the metal nanoparticle is modeled as a two-level quantum subsystem, with its creation and annihilation operators associated with transitions between ground and excited states.}
  \label{closed}
\end{figure}

In the specific case of a closed confined subsystem \(S_l\) of POP, it forms an integrated quantum system in which quantized TM polarized fields sourced in LSP oscillations and quasi-particle dynamics are intrinsically coupled (see Figure~\ref{closed}).

Each surface-cavity mode \( l \) is described using an anti-normally ordered Hamiltonian:

\begin{equation}
\hat{h}_l^{AN} = \hbar \omega_l \hat{a}_l \hat{a}_l^{\dagger},
\end{equation}

where \( \hat{a}_l^{\dagger} \) and \( \hat{a}_l \) are the photon creation and annihilation operators, respectively. These operators are mapped onto the raising and lowering operators \( \sigma_l^{+} = \ket{l}\bra{0} \) and \( \sigma_l^{-} = \ket{0}\bra{l} \), corresponding to transitions between the ground and excited many-electron states of the PQP. 
The equivalence \( \hat{a}_l^{\dagger} \leftrightarrow \sigma_l^{-} \) and \( \hat{a}_l \leftrightarrow \sigma_l^{+} \), reflects the process of photon emission corresponding to the transition \( \ket{l} \rightarrow \ket{0} \), while excitation corresponds to the reverse process.

The total Hamiltonian of the PQP system, spanning the Hilbert space of the uncoupled two-level modes, is: $\hat{H} = \sum_l \hat{h}_l$, where 
\begin{equation}
 \hat{h}_l =  \hbar \omega_l \sigma_l^{+} \sigma_l^{-} = \hbar \omega_l \ketbra{l}{l}.
\label{h}
\end{equation}

For each subsystem \( S_l \) with the ground state \( |0\rangle \) and  excited state \( |l\rangle \):
\begin{equation}
 \hat{h}_l |l\rangle = \hbar \omega_l |l\rangle, \qquad \hat{h}_l |0\rangle = 0.  \end{equation}

The ground state $|0\rangle$ represents the stationary, non-oscillatory state of the POP, that is, the state in the absence of collective electron excitations. In this sense, the zero-energy state of the plasmonic quasi-particle is an internal ground state of the many-electron system, defined with respect to the collective charge-density oscillations supported by the nanoparticle geometry. The excited states $|l\rangle$ correspond to discrete collective oscillation modes of the electrons with eigenenergies $\hbar \omega_l(R)$. 

In the formal bosonic quantization of plasmonic modes (e.g. \cite{scully1997quantum}), the full Hamiltonian includes a zero-point energy contribution of $\hbar\omega_l/2$. However, our description focuses on transitions between the excited states and the ground state, which involve energy differences of $\hbar \omega_l$. Consequently, the zero-point contribution cancels out and does not influence the decay dynamics or the self-quenching mechanism derived in this work. For this reason, the reduced Hamiltonian in {Eq.~(\ref{h})}, expressed in terms of the projection operators $\ketbra{l}{l}$, is entirely sufficient to describe the plasmonic quasi-particle energy structure relevant to emission and damping processes.

Such a self-consistent model provides a compact, physically grounded quantum framework for describing many-electron systems performing coherent collective motion. These systems can be viewed as micro- and macroscopic quantum structures (from tens to hundreds of nanometers in size), sharing characteristics related to the phenomenon of macroscopic quantum coherence demonstrated in superconducting circuits \cite{devoret1985measurements, martinis1985energy} and recently awarded by the Nobel Prize.

\subsection{Decay dynamics of Coherences and Populations in POP}

In realistic conditions, the plasmonic quasi-particle system \( S = \sum_l S_l \) interacts with its environment \( E \), forming an open quantum system. Environmental coupling introduces dissipation and decoherence, leading to energy loss and phase randomization. These effects can be approximately described within the Lindblad master equation formalism under the Markov approximation \cite{breuer2002theory}.

For a given subsystem \( S_l \), the evolution of the density matrix \( \rho^{S_l}(t) \) is given by:
\begin{equation}
\frac{d\rho^{S_l}}{dt} = -\frac{i}{\hbar} [\hat{h}_l, \rho^{S_l}] + D_l[\rho^{S_l}],
\end{equation}

where \( D_l[\rho^{S_l}] \) is the dissipator:
\begin{equation}
D_l[\rho^{S_l}] = -\frac{1}{2} \sum_{\alpha=r,nr} \left(
L_{\alpha,l}^{\dagger} L_{\alpha,l} \rho^{S_l} + 
\rho^{S_l} L_{\alpha,l}^{\dagger} L_{\alpha,l} 
- 2 L_{\alpha,l} \rho^{S_l} L_{\alpha,l}^{\dagger}
\right).
\end{equation}
The system evolves through two main decay channels:
radiative decay, associated with photon emission, modeled by \( L_{r,l} = \sqrt{2\Gamma_l^{reff}} \, \sigma_l^{-} \) and nonradiative decay, associated with internal losses (ohmic heating, electron–electron or electron–surface scattering), modeled by \( L_{nr,l} = \sqrt{2\Gamma_l^{nr}} \, \sigma_l^{-} \).
The radius dependence of the rates $\Gamma_l^{reff}(R)$ and $\Gamma_l^{nr}(R)$ with $\Gamma_l^{nr}= \Gamma^{nr}+\Delta_l$ follows from the classical modeling.

Solving the Lindblad equation for POP \cite{kolwas2019decay,kolwas2023optimization}
yields exponentially decaying populations and coherences: 

\begin{eqnarray}
\rho _{ll}(t) &=&\rho _{ll}(t_0)\exp \left( -2\Gamma _{l}t\right), \\
\rho _{00}(t) &=&\rho _{ll}(t_0)( 1-\exp ( -2\Gamma _{l}t)),  \\
\rho _{l0}(t) &=&\rho _{0l}(t_0) \exp
(( i\omega _{l}-\Gamma _{l})t)=
\rho _{0l} (t),  \label{DM}
\end{eqnarray}

where $\Gamma _{l}(R)$ are derived within the classical model (Eq.~(\ref{Gamma_l})). 

The population which are proportional to $\rho _{ll}(t)$ of the excited state decays as $e^{-2\Gamma_l t}$: while the oscillations of coherences (of the off-diagonal terms \( \rho_{l0}(t), \rho_{0l}(t) \)) decay twice as slowly, with a rate \( \Gamma_l \).
This behavior aligns with the general expectation that coherence decays more slowly than population, primarily due to its reliance on phase randomisation. In metal nanoparticles, the total rate \(\Gamma_l\), along with the radiative and non-radiative contributions, varies with nanoparticle size and is ultimately determined by the intrinsic electrodynamic properties of the metal nanoparticle. 

\subsection {Spontaneous Emission and Electromagnetic Environment}

Spontaneous emission in a free space, observed in quantum emitters like atoms or molecules, occurs due to vacuum electromagnetic fluctuations. These fluctuations enable transitions between energy states, with the emission rate determined by the spectral density of vacuum modes evaluated at the emitter’s transition frequency.

In a two-level atom, the spontaneous emission rate in vacuum is given by the Weisskopf–Wigner formula:

\begin{equation}
\Gamma_{21} = \frac{\omega_{21}^3 d_{12}^2}{3\pi\varepsilon_0\hbar c^3},
\end{equation}

where \( \omega_{21} \) is the transition frequency and \( d_{12} \) is the dipole matrix element. This expression is sometimes analogously applied to describe the radiative damping of dipolar LSP modes (e.g. \cite{moustafa2023bandwidth}).

When the emitter is placed in an inhomogeneous medium, such as an optical cavity, the spontaneous emission rate is modified by the available EM states of the surrounding environment, altering the vacuum field mode structure and the local density of states. As a result, the emission rate can be enhanced or suppressed compared to that in free space (e.g., \cite{martin2014quantum, novotny2012principles}. This fundamental sensitivity of spontaneous emission to the electromagnetic environment forms the basis of CQED  (e.g., \cite{maier2006effective, kristensen2012generalized, derom2012resonance, sauvan2013theory,pelton2015modified}) and underpins emission control in structured photonic media.

Cavity quantum electrodynamics (CQED) and its plasmonic counterpart (PQED) investigate how optical and plasmonic cavities affect spontaneous emission via the Purcell effect, which quantifies enhancement (or suppression) of emission due to increased LDOS near the emitter’s resonance \cite{haroche1989cavity, walther2006cavity, hugall2018plasmonic}.

Unlike free space, where EM modes form a continuum, a cavity supports discrete modes of frequency \( \omega_l \) and finite linewidth \( \Delta \omega_l \) of these modes. The quality factor of a cavity mode \( Q = \omega_l / \Delta \omega_l \), states how sharply localized the mode is in frequency. High-\( Q \) cavities concentrate the photonic density of states near the cavity's resonance frequency $\omega_{cav}$.

The Purcell factor $F_P$, expressing the enhancement of spontaneous emission relative to free space, is commonly expressed as:

\begin{equation}
F_P = \frac{3}{4\pi^2} \left( \frac{\lambda}{n} \right)^3 \frac{Q}{V_{eff}},
\end{equation}

where $\lambda$ is the emission wavelength, $n$ is the refractive index of the medium and $V_{eff}$ is the effective mode volume of the cavity.

The Purcell effect is directly related to the density of photonic states at the emitter's frequency $\omega_0$. Its derivation can be based on the arguments derived from Fermi's golden rule, which states that the transition rate for the emitter 
is proportional to the LDOS at the emitter’s resonance. 

For a resonant cavity, the LDOS can be approximated by a Lorentzian distribution:

\begin{equation}
\rho_{cav}(\omega) = \frac{1}{V_{eff}} \cdot \frac{\Delta \omega_{cav}}{(\Delta \omega_{cav})^2 + (\omega - \omega_{cav})^2}.
\label{ro_cav}
\end{equation}

 At resonance (\( \omega_{cav} =  \omega_0 \)), this simplifies to:
 
\begin{equation}
\rho_c(\omega_0) = \frac{Q}{\omega_0 V_{eff}}.
\label{ro_c}
\end{equation}

Thus, the Purcell factor relates to the ratio of cavity to free-space LDOS:

\begin{equation}
F_P = \frac{\Gamma_c}{\Gamma_f} = M \cdot \frac{\rho_c(\omega_0)}{\rho_f(\omega_0)} = \frac{1}{V_{eff}} \cdot \frac{Q}{\omega_0} \cdot \frac{\pi^2 c^3}{\omega_0^2},
\end{equation}

where \( M \) accounts for mode overlap. 
This expression highlights that maximizing the emission enhancement requires both a high $Q$ and a small mode volume $V_{eff}$.

For the conventional emitters in a plasmonic cavity, strong field confinement leads to very small \( V_{eff} \), enabling high Purcell enhancement despite low \( Q \). 
Plasmonic cavities of different geometries (e.g., bowtie, gap antenna) are essential tools in nanophotonics for squeezing light down to the nanoscale to unlock powerful light-matter interactions.
However, in proximity to metallic surfaces, nonradiative decay processes become predominant, leading to quenching of the emission and a concomitant decrease in the lifetime of the emitter (e.g. \cite{lakowicz2005radiative, anger2006enhancement, bharadwaj2006nanoplasmonic, guzatov2012plasmonic, olejnik2012plasmonic, rasskazov2018plasmon, kongsuwan2018suppressed, hugall2018plasmonic, hildebrandt2023plasmonic})

\section{Dual Functionality of Plasmonic Quasi-Particles}

In conventional cavity quantum electrodynamics (CQED) and plasmonic CQED (PCQED), spontaneous emission is typically studied in systems where an external cavity modifies the emission rate of an adjacent atom-like emitter. The cavity alters the local density of optical states (LDOS) available to the emitter, enhancing or suppressing spontaneous emission via the Purcell effect. In such systems, the emitter and the cavity are separate entities, and the cavity’s spectral properties remain unaffected by the emitter.

In contrast, the excited plasmonic nanoparticle plays a dual role, acting as a source of plasmonic cavity and the TM emitter embedded in a cavity of confined longitudinal near-field modes.   
The same physical structure allows the collective surface charge oscillations for a certain time in the femtosecond range (Fig.~\ref{fgr:1}a)) after the LSP excitation. These oscillations are a source of the locally confined longitudinal fields, which leak from the surface and decay as $1/r^2$.
The longitudinal evanescent field modes coupled to the oscillating surface charges do not radiate energy away. Instead, they act as a storage of reactive energy, while the tangential field components decay as $1/r$, carrying part of the energy to the far field region.
 
\begin{figure}
    \centering
    \includegraphics[width=8.0cm]{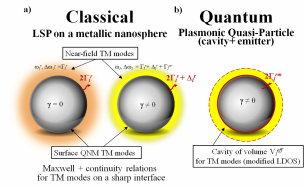}
    \caption{a) Classical picture of a metal nanoparticle supporting LSP resonant oscillations embedded in the near-field region, which contains the nonzero component of the electric field $E_l^r$ coupled to surface charge oscillations. The spectral characteristics of the interface and the near field are identical at and near resonance, but are modified in the collisionless electron regime. (b) Corresponding quantum picture: the near-field region is represented by the effective volume $V_l^{eff}$, forming a cavity with a modified LDOS. Plasmon energy is emitted into this self-generated cavity, which is inherently coupled to the emitter.}
    \label{fig:cavity}
\end{figure}

Accordingly, the interface and the adjacent near-field region can be conceptualized as the cavity of the effective volume $V_l^{eff}$ (Fig. \ref{fig:cavity}a)), with dimensions that scale with the size of the particle. 
Both the emitter and the cavity share the same spectral properties and are dynamically coupled.
This coupling is a defining feature of PQPs that has no counterpart in traditional atomic systems, leading to spontaneous emission dynamics fundamentally distinct from those in conventional atom-like or atom–cavity systems, where such mutual dependence does not exist.

Bridging the classical and quantum pictures, we adopt a self-consistent model where:

\begin{itemize}
    \item The plasmonic emitter is modeled as a sum of two-level quantum systems with excited state $\ket{l}$, ground state $\ket{0}$, and transition energy $\hbar \omega_l$. The decay of the exited state to the ground state takes place at the rate $\Gamma_l^{reff}$, which is the damping rate resulting from photon emission in the presence of electron collisions. 

    \item The plasmonic cavity is characterized by the effective volume $V_l^{eff}$ defining the evanescent near-field region, resonance frequency $\omega_l^c$, and total spectral width $\Gamma_l^c$ of the evanescent fields, resulting from both radiative and nonradiative contributions.
    
    \item The emitter and the cavity share identical spectral parameters, such that $\omega_l^c = \omega_l$ and $\Gamma_l^c = \Gamma_l=\Gamma_l^{reff}+\Gamma_l^{nr}$.
\end{itemize}

Therefore, nonradiative losses, such as electron scattering, affect not only the emitter's linewidth but also the photonic environment itself, by broadening the local density of optical states (LDOS) available for the emitter. This behavior fundamentally differs from the conventional Purcell effect, in which the cavity spectrum remains unaffected by the emitter.
   
In summary, the effective volume $V_l^{eff}$ provides a geometric and physical link between the classical near-field confinement and the quantum LDOS into which emission occurs (Fig. \ref{fig:cavity}). This bridge enables a consistent formulation of spontaneous emission rates for plasmonic quasi-particles based on Fermi’s golden rule, incorporating both the emitter and the structured EM environment as a single, self-consistent system.

\subsection{Quality Factor of Plasmonic Cavity }

The quality factor of the plasmonic cavity $Q_l(R)$, \cite{derom2012resonance,kristensen2014modes,stockman2018roadmap,kolwas2020impact} defined as the number of oscillations required for the energy of a freely oscillating system to fall off to $e^{-2\pi}$ of its original energy, is a function of the radius $R$ of MNP: 
  
\begin{equation}
 Q_l(R)= \hbar\omega_l^c(R)/\Delta E_l(R) = \omega_l^c(R)/\Gamma_l^c(R), \label{QF}  
\end{equation}

where $\omega_l^c$ is the resonance frequency of the mode and $\Delta\omega_l^{c}=\Gamma_l^c$ is the total linewidth of the cavity mode due to both radiative and nonradiative processes.

As the surface TM fields and the concentrated fields around the plasmonic emitter possess the same spectral parameters, solutions of the dispersion relation (Eq.~(\ref{DR})) yield:

\begin{equation}
    \Gamma_l(R)=\Gamma_l^c(R), \label{equiv}
\end{equation}

where $\Gamma_l^c$ is composed of the radiative and nonratiative  contributions (Eq~(\ref{Gamma_l})).

The quality factor thus encapsulates both the intrinsic resonant confinement of the EM field (via $\omega_l^c$) and the loss mechanisms that limit the ability of the cavity to store the total energy of LSP excitation. It plays a central role in determining LDOS, which in turn controls the emission dynamics of the plasmonic emitter.

\subsection{Density of States Available for Plasmonic Emitter }

The LDOS of plasmonic cavities (including those created by a PQP emitter) depends on the spectral and spatial characteristics of the LSP; therefore, it differs from the LDOS of freely decaying atomic systems in which the photonic environment is unstructured, but also from the LDOS of optical cavities (Eq.~(\ref{ro_cav})).
The spectral identity between emitter and cavity makes the plasmonic emitter fundamentally different from standard emitters studied in CQED and PCQED.

The LDOS quantifies the number of electromagnetic modes available for photon emission per unit volume and frequency. Therefore, in the case of a single-mode plasmonic cavity characterised by an effective volume $V_l^{eff}$ and resonance frequency $\omega_l$, the LDOS (Eqs.~(\ref{ro_cav}) of PQP's created cavity can be expressed as: 

\begin{equation}
\rho^{p}(\omega_l) 
= \frac{1}{V_l^{eff} \Delta\omega_l^{c}}
= \frac{Q_l}{\omega_l V_l^{eff}},
\label{ro_p}
\end{equation}
where $\Delta\omega_l^{c}$ is the full width at half maximum (FWHM) of the cavity resonance, linked to the total loss rate $\Delta\omega_l^{c}= \Gamma_l^c = \Gamma_l$  (Eqs.~(\ref{Gamma_l}), (\ref{equiv}) of the mode and $Q_l$ is the quality factor.

Consequently, the density of states $\rho^{p}$ that is available for the emitter is affected by intrinsic material losses, such as ohmic attenuation and surface scattering (Eq.~\ref{gamma}). 

Conversely, traditional CQED and PCQED systems use external, independent resonators. Within the PQP framework, the correspondence between the emitter and cavity parameters necessitates a novel, self-consistent formulation of the spontaneous emission decay from the confined plasmonic systems.

\subsection{Extension of Fermi’s Golden Rule to POP Systems} 

Fermi’s Golden Rule (FGR) provides a foundational quantum framework for calculating transition rates in systems weakly coupled to a continuum of final states. For spontaneous emission, it is typically given by:

\begin{equation}
\Gamma_{if} = \frac{2\pi}{\hbar} |\langle f | \hat{H}' | i \rangle|^2 \rho(E_f),
\label{FGR_standard}
\end{equation}
where $\ket{i}$ and $\ket{f}$ are the initial and final quantum states, $\hat{H}'$ is the interaction Hamiltonian, and $\rho(E_f)$ is the density of final states at energy $E_f$. This expression assumes an external, EM environment independent of the emitter.

By contrast, a plasmonic quasi-particle (PQP) constitutes a self-consistent, strongly coupled system in which both the emitter and the EM environment arise from the same underlying physical structure and are characterized by identical spectral parameters. In this case, radiation is emitted into a spatially structured, near-field EM environment that remains dynamically coupled to the emitter.

To account for this, we extend FGR by incorporating the cavity LDOS (Eq.~\ref{ro_p})): 
\begin{equation}
\Gamma_l^p = \frac{2\pi V_l}{\hbar} |\langle l | \hat{V} | 0 \rangle|^2 \rho^{p}(\hbar \omega_l) = \frac{2\pi V_l}{\hbar^2}  \frac{|\langle l | \hat{V} | 0 \rangle|^2}{V_l^{eff} \Gamma_l^c},
\label{FGR_extended}
\end{equation}

where $\Gamma_l^p = 2\Gamma_l^{r\,eff}$ denotes the spontaneous emission rate of the plasmonic quasi-particle, and $\Gamma_l^c = \Gamma_l^{r\,eff} + \Delta_l+\Gamma^{nr}$ is the total linewidth of the LSP resonance (Eq.~\ref{Gamma_l}) and therefore that of the PQP cavity (Eq.~\ref{equiv}), with parameters adjusted to the damping rates of the emitter. 
    The reduction term $\Delta_l$, which arises from the nonorthogonality of the eigenstates of the open system has been incorporated in $\Gamma_l^{nr}$.
The matrix element $|\langle l | \hat{V} | 0 \rangle|^2$ is renormalized by a mode confinement factor $\beta_l = V_l / V_l^{eff}$.

The collisionless limit ($\Gamma_l^{nr} = 0$), Eq.~(\ref{FGR_extended}), allows us to connect the radiative rate $\Gamma_l^r$ to the perturbation strength between the final and initial states:

\begin{equation}
2\Gamma_l^r = \frac{2\pi V_l}{\hbar^2} |\langle l | \hat{V} | 0 \rangle|^2 \frac{1}{V_l^{eff} \Gamma_l^{c\,r}},
\end{equation}

which allows us to express the interaction strength as:
\begin{equation}
\frac{2\pi}{\hbar^2} |\langle l | \hat{V} | 0 \rangle|^2 = \frac{2 \Gamma_l^r \Gamma_l^{c\,r} V_l^{eff}}{V_l}.
 \label{eq:interaction}
 \end{equation}

Substituting this into Eq.~(\ref{FGR_extended}) for the general lossy case gives:
\begin{equation}
\Gamma_l^p =2 \Gamma_l^{reff} = \frac{2 \Gamma_l^r \Gamma_l^{c\,r}}{\Gamma_l^c} = \frac{2 (\Gamma_l^r)^2}{\Gamma_l^{reff} + \Gamma_l^{nr}}. \label{G}
\end{equation}

Solving Eq.~(\ref{G}) self-consistently yields the effective radiative decay rate of the plasmonic quasi-particle. The resulting analytical expression constitutes the central result of the present model, as it explicitly reveals the feedback between intrinsic nonradiative dissipation and the radiative emission process:
\begin{equation}
\Gamma_l^{reff} = \frac{1}{2} \left( \sqrt{4(\Gamma_l^r)^2 + (\Gamma_l^{nr})^2} - \Gamma_l^{nr} \right) \leq \Gamma_l^r.
\label{Gamma_em_final}
\end{equation}

In the limiting case $\Gamma_l^{nr} \rightarrow 0$, Eq.~(\ref{Gamma_em_final}) correctly reduces to $\Gamma_l^{r\,eff} = \Gamma_l^r$, recovering the conventional radiative decay rate. 

Expression \ref{Gamma_em_final} reveals a self-quenching effect, in which a nonradiative decay channel attenuates radiative emission as a consequence of the dynamic coupling between the emitter and the cavity. This mechanism is fundamentally different from classical Purcell suppression, since it arises from the dissipative characteristics of the emitter’s immediate self-created EM environment. 

Nonetheless, it can be viewed as analogous to fluorescence quenching in conventional PQED architectures, which occurs when an atomic or molecular emitter is placed in close proximity to a plasmonic nanoparticle \cite{lakowicz2005radiative, anger2006enhancement, bharadwaj2006nanoplasmonic, guzatov2012plasmonic, olejnik2012plasmonic, rasskazov2018plasmon, kongsuwan2018suppressed, hugall2018plasmonic, hildebrandt2023plasmonic}.

In limiting cases:
\begin{center}
    If \( \Gamma_l^{nr} \ll \Gamma_l^r \), then \( \Gamma_l^{reff} \approx \Gamma_l^r \);\\
    If \( \Gamma_l^{nr} \gg \Gamma_l^r \), then \( \Gamma_l^{reff} \ll \Gamma_l^r \).
\end{center}
    
In general, the total damping rate of the PQP becomes:

\begin{equation}
\Gamma_l = \Gamma_l^{r\,eff}+ \Gamma_l^{nr} \leq \Gamma_l^r(R) + \Gamma_l^{nr}.
\label{Gamma_l_quan}
\end{equation}

The structure of Eq.~(\ref{Gamma_em_final}) explicitly illustrates the non-additive character of the decay channels. The nonradiative, collisional losses in metal NPs do not merely contribute linearly to the LSP damping. Instead, they can dynamically suppress the total decay rate within certain regions of the parameter space.

Although the self-quenching effect described by E{q.~(\ref{Gamma_em_final})} supports the central claim of this paper that plasmon damping cannot in general be treated as a simple sum of independent decay channels ({Eq.~(\ref{Gamma_l})} and {Fig.~\ref{fgr:reduction}}), a quantitative evaluation of the effect versus size requires separating the radius dependence of the radiative contribution $\Gamma_l^r(R)$ from the total damping rate $\Gamma_l(R)$. Such a decomposition was not performed in {Ref~\cite{kolwas2013damping}}. In addition, a complete parametrization of interband transitions, which contribute to nonradiative damping rates $\gamma_l^{nr}(R)$ in some size ranges, remains incomplete. A systematic study addressing these aspects is currently in progress and will be reported separately.

\subsection{Comments on Some Experimental Results in Nanorods}

 Experimental investigations of localized surface plasmon (LSP) damping in noble-metal nanoparticles conventionally represent the total damping rate as the linear superposition of independent radiative and nonradiative contributions. Nonetheless, several studies (e.g. \cite{sonnichsen2002drastic, moustafa2023bandwidth, kolwas2013damping, ginzburg2012non} have reported anomalous behavior that departs from the standard assumptions in specific systems and experimental conditions.

In \cite{sonnichsen2002drastic}, the authors study plasmon damping rates in single gold nanoparticles using dark-field spectroscopy to record light-scattering spectra as a function of particle size and shape. The study demonstrates that in the spherical NPs, the homogeneous broadening (dephasing rate) of the dipole plasmon resonance increases with particle size due to enhanced radiation damping, as expected. In contrast, in nanorods, the total dephasing rate of a low-radiative LSP dipole resonance decreases drastically with increasing length. The authors attribute this unexpected effect to suppressed interband damping, which contributes to nonradiative damping. Their supporting quasistatic calculations neglect radiation damping.

The results for the dipole mode of nanospheres remain in accordance with our expectations in the studied range of sizes $2R=$ 20,..., 150 nm, as illustrated in Fig.~\ref{fgr:reduction}b) (black line).

Nanorods, in contrast, offer more favorable conditions for observing the coupling between the radiative and nonradiative channels, thereby reducing the total damping rate. The long-axis mode behaves nearly as an ideal dipolar emitter with low radiative losses, even for larger sizes. Therefore, the condition  $\Gamma_1^{nr} \gtrsim \Gamma_1^r$ is fulfilled over a wider range of sizes than in the case of spherical MNPs. According to {Eqs.~(\ref{Gamma_em_final})} and  {Eqs.~(\ref{Gamma_l_quan})} in such a range of parameters, one can expect suppression of the radiative (and the total damping) rate with the increasing NP size.

More direct evidence of the coupling between radiative and nonradiative channels was provided by recent TR-PEEM (Time-Resolved PhotoEmission Electron Microscopy) experiments on gold nanorods \cite{qin2023coaction}. The authors measured lifetimes of bright (superradiant) and dark (subradiant) modes as a function of nanorod length. Surprisingly, bright modes were found to have lifetimes equal to or longer than those of subradiant modes, contradicting the expectation that radiative losses shorten the lifetimes of bright modes: $ \Gamma^{super}(\Gamma^r,\Gamma^{nr}) \leq\Gamma^{subr}(\Gamma^{nr})$, despite the presence of radiative damping in the superradiant case. This confirms the co-dependence of decay channels Eq.~\ref{Gamma_em_final}).
   
Such behavior provides qualitative support for our prediction, evidencing the non-additive character of the decay channels that results in a reduction of the experimentally observed total damping rates in metallic nanoparticles. In plasmonic systems, collisional, heat-generating losses can dynamically suppress the overall decay rates within specific regions of parameter space.

\section{Conclusions}

This work introduces a unified, quantum-informed framework in which decaying localized surface plasmon (LSP) excitations are treated as plasmonic quasiparticles: quantized, coherent many-electron emitters that radiate energy into their own structured electromagnetic near-field environment of longitudinal fields. In this description, the plasmonic excitation simultaneously acts as both the emitter and the source of the confined electromagnetic modes forming the effective cavity. Unlike conventional cavity quantum electrodynamics (CQED), where the emitter and cavity are treated as independent subsystems, the present framework considers them as dynamically coupled parts of the same physical system. The implications of this intrinsic coupling for the damping dynamics of coherent electron oscillations constitute the central conceptual aspect of the present study.

An intrinsic suppression of the total damping rate of radiative modes due to the self-quenching effect is predicted. Unlike in the traditional Purcell effect, this modification of the damping rate does not arise from an externally altered EM environment. Rather, it results from the loss of a portion of the excitation energy via collisional, ohmic processes in a metal nanoparticle, leading to heat generation. Consequently, the emission characteristics of the plasmonic system are modified self-consistently: the effective LDOS, which depends on the quality factor of the self-created plasmonic cavity, is itself influenced by nonradiative losses, introducing a strict co-dependence of the radiative and nonradiative decay channels.

The original contribution of this work lies in the formulation of a self-consistent extension of Fermi’s Golden Rule for plasmonic quasi-particles, in which the radiative decay rate depends explicitly on the total cavity linewidth entering the LDOS. This feedback mechanism leads to an analytical expression for the effective radiative rate (Eq.~29), revealing a non-additive coupling between decay channels and predicting a size- and mode-dependent self-quenching effect. Unlike conventional phenomenological approaches that decompose the total linewidth into independent radiative and nonradiative terms, the present framework demonstrates that intrinsic dissipation modifies the emission process itself through the shared spectral parameters of the emitter and its self-generated cavity. Thus, the conventional description of LSP decay as a simple sum of independent radiative and nonradiative channels fails to fully capture the dynamics of plasmonic systems in which the emitter and cavity are intrinsically coupled. 

These results are consistent with the classical description of transient LSP decay within Maxwell electrodynamics and help explain experimental observations of anomalously long lifetimes in nanorods \cite{sonnichsen2002drastic, qin2023coaction}. 

Classically derived cavity parameters, obtained from the dispersion relation of quasi-normal transverse-magnetic (TM) multipolar modes—including surface scattering and retardation effects—form the physical basis of our model. This framework remains valid over a broad size range, from individual nanometer-scale structures up to several hundreds of nanometers, and extends well beyond the dipole approximation. The reinterpretation of these classical parameters within a self-consistent quantum framework provides a physically transparent bridge between the Maxwell and quantum descriptions of LSP dynamics.

Beyond its implications for emission control, the model highlights the bosonic nature of PQPs: coherent excitations of many-electron systems that behave collectively as a single quantized state. These systems can be viewed as micro- and macroscopic quantum structures sharing characteristics related to macroscopic quantum coherence. 
These localized surface-bound bosonic modes form without nonlinear thresholds and can exist at room temperature. Unlike traditional Bose–Einstein condensates (BECs), which require ultralow temperatures and massive particles, PQPs provide a platform for studying room-temperature bosonic phenomena in three-dimensional systems. 
This insight aligns with recent demonstrations of polariton and quasiparticle condensation under ambient conditions~\cite{lagoudakis2010coherent, kkedziora2024predesigned}, further supporting the broader relevance of the PQP framework.

The self-consistent, quantum-local approach presented here contributes to a deeper understanding of decay processes in coherent many-electron micro- and macro-systems. Ultimately, it may serve in the development of decay engineering strategies, controlled light–matter interactions, and future quantum technologies based on highly dissipative yet coherent plasmonic systems.

\begin{acknowledgement}

This research was funded in whole or in part by the National Science Centre, Poland, grant 2021/41/B/ST3/00069. For the purpose of Open Access, the authors have applied a CC-BY public copyright license to any Author Accepted Manuscript (AAM) version arising from this submission. 

\end{acknowledgement}
\bibliography{Density_matrix}

\end{document}